\documentclass{article}

\usepackage{Definitions/arxiv}

\usepackage[utf8]{inputenc} 
\usepackage[T1]{fontenc}    
\usepackage{hyperref}       
\usepackage{url}            
\usepackage{booktabs}       
\usepackage{amsfonts}       
\usepackage{nicefrac}       
\usepackage{microtype}      
\usepackage{lipsum}
\usepackage{graphicx}
\usepackage{amsmath}
\usepackage{amsthm}
\usepackage{bm}
\usepackage{cases}
\newtheorem*{remark}{Remark}
\usepackage{enumitem}
\usepackage{graphicx}
\usepackage{subcaption}
\usepackage{float}
\usepackage{multirow}
\graphicspath{ {figures/} }

\newlist{steps}{enumerate}{1}
\setlist[steps, 1]{wide=0pt, leftmargin=\parindent, label=Step \arabic*:, font=\bfseries}
\newcommand\sbullet[1][.5]{\mathbin{\vcenter{\hbox{\scalebox{#1}{$\bullet$}}}}}

\title{Quantifying the advantage of vector
over scalar magnetic sensor networks for undersea surveillance}
\author{
 Wenchao Li \\
  School of Science\\
  RMIT University\\
  Melbourne, Australia\\
  \texttt{lwenchao23@gmail.com} \\
   \And
 Xuezhi Wang \\
  School of Science\\
  RMIT University\\
  Melbourne, Australia\\
  \texttt{xuezhi.wang@rmit.edu.au} \\
  \And
 Qiang Sun \\
  School of Science\\
  RMIT University\\
  Melbourne, Australia\\
  \texttt{qiang.sun@rmit.edu.au} \\
  \And
    Allison N. Kealy \\
  Innovative Planet Research Institute\\
  Swinburne University of Technology\\
  Melbourne, Australia \\
  \texttt{akealy@swin.edu.au} \\ 
  \And
   Andrew D. Greentree \\
  School of Science\\
  RMIT University\\
  Melbourne, Australia\\
  \texttt{andrew.greentree@rmit.edu.au}
}

\begin{document}
\maketitle
\begin{abstract}
Magnetic monitoring of maritime environments is an important problem for monitoring and optimising shipping, as well as national security.  New developments in compact, fibre-coupled quantum magnetometers have led to the opportunity to critically evaluate how best to create such a sensor network.  Here we explore various magnetic sensor network architectures for target identification.  Our modelling compares networks of scalar vs vector magnetometers.  We implement an unscented Kalman filter approach to perform target tracking, and we find that vector networks provide a significant improvement in target tracking, specifically tracking accuracy and resilience compared with scalar networks.  
\end{abstract}

\section{Introduction}

Monitoring of maritime traffic is important for a range of applications, including managing congestion, identification of threats, and national security.  As most maritime platforms have some magnetic signature, magnetometry is an important sensor modality for such surveillance~\cite{feng2020magmonitor,wahlstrom2013magnetometer}.

Magnetometers for maritime surveillance can either be mobile platforms, for example aerial drones~\cite{seidel2023underwater}  uncrewed undersea drones~\cite{XYN+2016,oehler2024processing,YZZ+2025}, or towed arrays \cite{EBG2008}.  Alternatively, for specific regions of interest, magnetic trip lines \cite{SFZ+2022} or arrays \cite{DMR18} can be considered.  A magnetic trip line is a line or network of magnetometers that might be placed on the sea bed.  The magnetic signals detected are then correlated to determine when an object passes over the sensors.  This is the focus of the current work.  Because the magnetic field from metallic objects decreases as the cube of the distance, magnetic sensor networks are most useful for monitoring in relatively shallow water. 

Deployment of traditional magnetometers for long-term undersea monitoring is relatively costly due to the size and power requirements of magnetometers with sufficient sensitivity. 
Quantum approaches provide a new context to explore undersea magnetometry due to their reduced size weight and power for the same sensitivity as traditional classical magnetometers.  In particular, optically pumped magnetometers \cite{BR2007} provide outstanding scalar magnetometry, with sensitivity reaching down to tens of femtotesla per square root Herz~\cite{aleksandrov2009modern, oelsner2022integrated}.  New generations of diamond magnetometers also provide outstanding sensitivity including vector sensing~\cite{graham2025road,10.1093/nsr/nwae478}, with record sensitivity of around $500 \text{fT}/\sqrt{\text{Hz}}$~\cite{BSA+2024,SLR+2025}. Diamond in fibre approaches promise robust housing and low size, weight and power (SWaP)~\cite{HGE+2011,RSJ+2018,BHS+2020,patel2020subnanotesla,FMS+2022}. Though the sensitivity is not as high as conventional diamond magnetomers, diamond in fibre can still achieve 30~pT/$\sqrt{\text{Hz}}$~\cite{graham2023fiber}.  These levels of sensitivity should be compared with the expected noise floors in marine environments, which are typically around the hundreds of pT level \cite{AGT+2009,GLD+2020}, depending on the conditions.  Such noise can also be mitigated by emerging de-noising techniques \cite{GDL+2016} although we will not consider such details here.

Here we show a comparison of magnetometer sensing networks with a particular emphasis on sensitivity at current or near-future sensitivity for quantum sensors.  We compare the efficiency of target tracking for single path and periodic targets, quantifying the improvement through use of vector vs scalar sensors.  We find that vector sensors significantly outperform scalar sensors, by more than a factor of three.  These results show that a sparser vector network can outperform a more dense scalar network, which has implications for the robustness and cost of implementing and operating such magnetometer networks for persistent monitoring applications.

This manuscript is organised as follows: We first introduce our measurement and system models; we then describe our centralised unscented Kalman filter; lastly we introduce two measurement scenarios.

\section{Problem formulation}

\begin{figure}[tb!]
    \centering
    \includegraphics[width=1\textwidth,trim={3.5cm 0 0 0},clip]{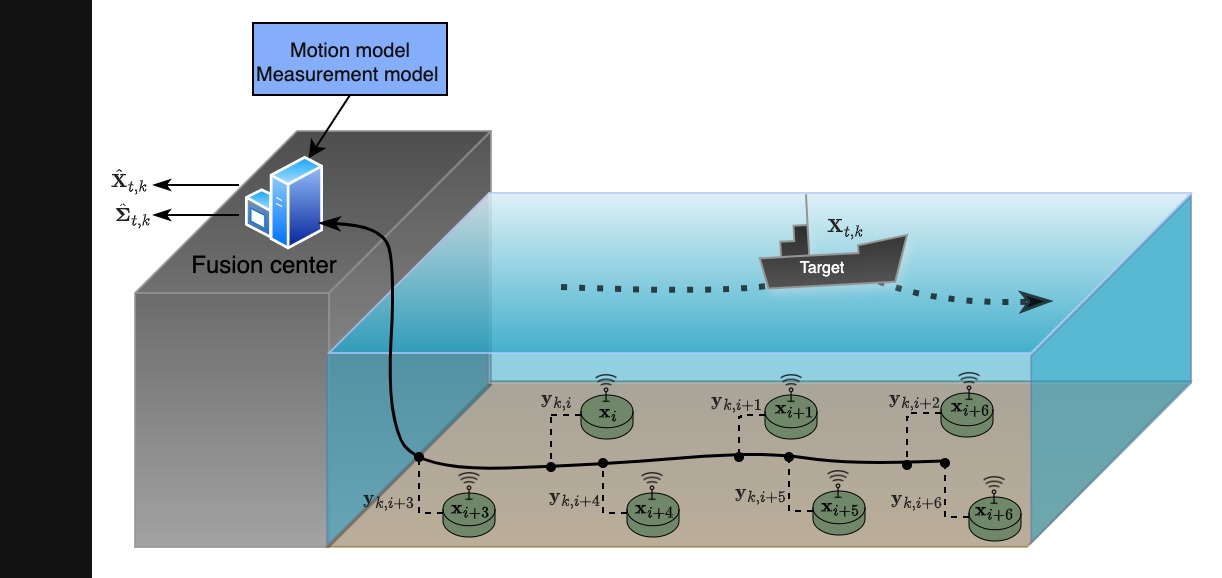}
    \caption{Estimating the target's true location $\mathbf{X}_{t,k}$ at time $k$ via centralized fusion and sensor network composing with $n$ sensors.  These sensors are located at $\mathbf{x}_{i}$ and can provide magnetic scalar measurement or magnetic field measurement to the fusion center for tracking target. 
    }
    \label{fig:scheme}
\end{figure}

In general, targets of interest for tracking will have a non-trivial magnetic signature.  Nevertheless, when the distance between target and sensor is three times longer than the size of the target, we can approximate the target as a single magnetic dipole~\cite{SG2000}.  A schematic of our system is shown in Fig.~\ref{fig:scheme}.

At time $k=1,2,3,\cdots$, the location of target dipole is  $\mathbf{x}_{t,k} = [x_{t,k},y_{t,k},z_{t,k}]^T$, where the subscript $t$ in $\mathbf{x}_{t,k}$ indicates the target. The motion of the target is described by the state 
\begin{align}
    \mathbf{X}_{t,k}= \left[\mathbf{x}_{t,k}^T, \dot{\mathbf{x}}_{t,k}^T,\ddot{\mathbf{x}}_{t,k}^T\right]^T \label{complete_state}
\end{align}
where $\dot{\mathbf{x}}_{t,k}$ and $\ddot{\mathbf{x}}_{t,k}$ are the target's true velocity vector and acceleration vector respectively.

The magnetic field from the target dipole as measured at the $i$-th static sensor at location $\mathbf{x}_i = [x_i,y_i,z_i]^T$, $i=1,2,\cdots,n$, is~\cite{mcfee1990locating,wiegert2005generalized}
\begin{align}
\mathbf{B}(\bar{\mathbf{x}}_{i,k})=&\frac{\mu_0}{4\pi}\left(\frac{3(\mathbf{M}\cdot\bar{\mathbf{x}}_{i,k})\bar{\mathbf{x}}_{i,k}}{r_{i,k}^5}-\frac{\mathbf{M}}{r_{i,k}^3}\right)\notag\\
=&\frac{\mu_0}{4\pi r^5_{i,k}}\left(3\bar{\mathbf{x}}_{i,k}\bar{\mathbf{x}}_{i,k}^T-r^2_{i,k}\mathbf{I}_3\right)\mathbf{M}
\end{align}
 where $\bar{\mathbf{x}}_{i,k} = [\bar{x}_{i,k},\;\bar{y}_{i,k},\;\bar{z}_{i,k}]^T =\mathbf{x}_i- \mathbf{x}_{t,k}$ is the vector between the $i^{\text{th}}$ sensor and the target at time $k$, $r_{i,k} = ||\bar{\mathbf{x}}_{i,k}||_2$ is the sensor-target distance at time $k$, and $\mathbf{I}_3$ is the 3-dimensional identity matrix. $\mathbf{M}=[M_{x}, M_{y}, M_{z}]^T$ is the magnetic moment of the target dipole and $\mu_{0}$ is the permeability of the medium, assumed to be sea water.

We treat two different measurement cases as follows.
\begin{itemize}

\item \textbf{Scalar measurement}:

The case of scalar measurements corresponds to sensing via scalar magnetometers, for example optically pumped magnetometers~\cite{gerginov2017pulsed}.  In this case, at time step $k$, the $i$-th sensor measures the norm of $\mathbf{B}(\bar{\mathbf{x}}_{i,k})$, i.e., $\|\mathbf{B}(\bar{\mathbf{x}}_{i,k})\|_2$. The measurement model becomes
\begin{align}
y_{i,k} = \|\mathbf{B}(\bar{\mathbf{x}}_{i,k})\|_2+{\omega}_{i,k}\label{meas_model_scalar}
\end{align}
where ${\omega}_{i,k}\sim\mathcal{N}(0,\sigma^2)$ is noise term with zero mean and variance $\sigma^2$. 

\item \textbf{Vector measurement}:

The vector case corresponds to an array of vector magnetometers, such as are realised by nitrogen-vacancy diamond magnetometers~\cite{manian2025nitrogen}.  The vector measurement model for sensor $i$ at time step $k$ is
\begin{align}
\mathbf{y}_{i,k} =
\mathbf{B}(\bar{\mathbf{x}}_{i,k})+\bm{\omega}_{i,k}\label{meas_model_vector}
\end{align}
where $\bm{\omega}_{i,k}\sim\mathcal{N}([0,0,0]^T,\bm{\Sigma})$ is the noise term treated as Gaussian distributed with with zero mean, and variance matrix $\bm{\Sigma}=\sigma^2\mathbf{I}_3$, where $\sigma$ is the noise variance.
\end{itemize}

\begin{remark}
As scalar and vector sensors typically use different technologies, there is no guarantee that the noise levels will be comparable. However, for simplicity in our analysis and to provide fair comparisons between implementations, we assume that the variance matrix $\Sigma$ in \eqref{meas_model_vector} is a diagonal matrix with identical diagonal elements that is equal to 
the noise variance in the scalar case~\eqref{meas_model_scalar}, i.e. $\sigma^2$.

\end{remark}

Considering the state of target at time $k$, $\mathbf{X}_{t,k}$ defined in~\eqref{complete_state}, the state-space system with scalar measurement or vector measurement can be modeled as
\begin{subnumcases}
{} \mathbf{X}_{t,k} = f(\mathbf{X}_{t,k-1},\bm{u}_{k})\label{motion}\\
 y_{i,k} = \|\mathbf{B}(\bar{\mathbf{x}}_{i,k})\|_2+{w}_{i,k}\;\;\text{or} \;\;\mathbf{y}_{i,k} =
\mathbf{B}(\bar{\mathbf{x}}_{i,k})+\bm{\omega}_{i,k}\label{mea}
\end{subnumcases}
where \eqref{motion} describes the motion of the vehicle, \eqref{mea} is the measurement model, $\bm{u}_{k}$ is the process noise which is assumed to be Gaussian distributed with zero mean and covariance $\mathbf{Q}_k$ and $\bar{\mathbf{x}}_{i,k} = \mathbf{H}\mathbf{X}_{t,k}-\mathbf{s}_i=\mathbf{x}_{t,k}-\mathbf{s}_i$ where $\mathbf{H}$ is given by
\begin{align}
    \mathbf{H} = \begin{pmatrix}
        \mathbf{I}_3\;|\;\mathbf{0}_3\;|\;\mathbf{0}_3
    \end{pmatrix},
\end{align}
and is used to extract the target position from the full target state, where $\mathbf{0}_3$ is the 3-dimensional zero matrix.

The target tracking problem with the model~\eqref{motion} and~\eqref{mea} is to find the posterior density of target state over time based on the collection of magnetic field measurements from the sensor network, i.e., $p(\mathbf{X}_{t,k}|y_{i,j}, i=1\cdots,n \text{ and }j = 1,\cdots,k)$. In the following section, we solve this problem using a Bayesian centralized data fusion, where unscented Kalman filter (UKF) is used for nonlinear state estimation of the target. The estimators are established for both the scalar and vector measurement models.

\section{Centralized Unscented Kalman Filter}

Our system considers a sensor array in a network.  To perform target tracking, we assume that all sensors report their measurements to a common center for processing and tracking, i.e. centralised tracking. Since the measurement is highly nonlinear as seen in~\eqref{mea}, a nonlinear filter, such as extended Kalman filter (EKF), unscented Kalman filter (UKF), or particle filter, should be applied. Although a particle filter can provide better performance in general, it requires significant computational resource. On the other hand, the UKF is widely considered superior to the EKF~\cite{hao2007comparison, hostettler2014vehicle} with possibly slightly higher computational cost. Furthermore, UKF effectively evaluates both Jacobian and Hessian precisely through its sigma point propagation, without calculating the analytic Jacobian matrix as in EKF~\cite{wan2000unscented}. Therefore, we choose to employ UKF as the tracker. 

The fusion center receives all the measurements available from the sensors in the network. Therefore, in the UKF, a stacked measurement vector, $\mathbf{Y}^s_k$ for scalar measurement network or $\mathbf{Y}^v_k$ for vector measurement network, should be used and can be defined as follows.
\begin{align}
    \mathbf{Y}^s_k=&\left[\|\mathbf{B}(\bar{\mathbf{x}}_{1,k})\|_2,\|\mathbf{B}(\bar{\mathbf{x}}_{2,k})\|_2,\cdots,\|\mathbf{B}(\bar{\mathbf{x}}_{n,k})\|_2\right]^T + \mathbf{W}^s_{k}\label{stacked_measurement_s}\\
    \text{or,      } \mathbf{Y}^v_k=&\left[\mathbf{B}(\bar{\mathbf{x}}_{1,k})^T,\mathbf{B}(\bar{\mathbf{x}}_{2,k})^T ,\cdots,\mathbf{B}(\bar{\mathbf{x}}_{n,k})^T \right]^T+ \mathbf{W}^v_{k}\label{stacked_measurement_v}
\end{align}
where $\mathbf{Y}^s_k\in \mathbb{R}^{n\times1}$, $\mathbf{Y}^v_k\in \mathbb{R}^{3n\times1}$, $$\mathbf{W}^s_{k}=\left[\omega_{1,k},\omega_{2,k},\cdots,\omega_{n,k}\right]^T\sim\mathcal{N}(\mathbf{0}_n, \sigma^2\mathbf{I}_n)$$ and $$\mathbf{W}^v_{k}=\left[\bm{\omega}_{1,k}^T,\bm{\omega}_{2,k}^T,\cdots,\bm{\omega}_{n,k}^T\right]^T\sim\mathcal{N}(\mathbf{0}_{3n}, \sigma^2\mathbf{I}_{3n}).$$

The details of UKF is given as follows. For the $9$-dimensional state $\mathbf{X}_{k}$ defined in~\eqref{complete_state} with mean $\hat{\mathbf{X}}_{k|k}$ and variance $\bm{\Sigma}_{k|k}$ the unscented transform involves $2N+1$, where $N=9$, sigma points with weights 
\begin{subnumcases}
{a_i = } \frac{\kappa}{N+\kappa}\;\;\;\;\;\;i=0\\
  \frac{1}{2(N+\kappa)}\;\;\text{otherwise}
\end{subnumcases}
where $\kappa\in\mathbb{R}^+$.

Suppose that, the estimate of $\mathbf{X}_k$ at time $k-1$ is $\hat{\mathbf{X}}_{k-1|k-1}$ with covariance $\bm{\Sigma}_{k-1|k-1}$, then at time $k$, the recursion is as follows
\begin{steps}
    \item Calculate sigma points $\hat{\mathbf{X}}^j_{k-1|k-1}$, $j=0,\cdots,2N$ by
    \begin{align}
    \hat{\mathbf{X}}^j_{k-1|k-1} &=\hat{\mathbf{X}}_{k-1|k-1} +s\sqrt{N+\kappa}L_i\label{sigma}
    \end{align}
    where $s=0$ for $j=0$, $s=1$ for $j=1,\cdots,N$, $s=-1$ for $j=N+1,\cdots,2N$ and  $L_j$ is the $j$-th column of $\mathbf{L}$ with $\mathbf{L}\mathbf{U}=\bm{\Sigma}_{k-1|k-1}$ be the LU decomposition of covariance $\bm{\Sigma}_{k-1|k-1}$.

    \item Predict the state and covariance by
    \begin{align}
        \hat{\mathbf{X}}_{k|k-1}=&\sum_{j=0}^{2N}a_jf(\hat{\mathbf{X}}_{k-1|k-1})\\
        \bm{\Sigma}_{k|k-1}=&\sum_{j=0}^{2N}a_j(\hat{\mathbf{X}}^j_{k-1|k-1}-\hat{\mathbf{X}}_{k|k-1})(\hat{\mathbf{X}}^j_{k-1|k-1}-\hat{\mathbf{X}}_{k|k-1})^T+\mathbf{Q}_K.
    \end{align}
    \item The measurement update. The sigma points of location difference between target location and the $i$-th sensor can be calculated by $\hat{\bar{\mathbf{x}}}^{j}_{i, k-1|k-1} = \mathbf{H}\hat{\mathbf{X}}^j_{k-1|k-1}-\mathbf{s}_i$, then the measurement prediction is
    \begin{align}
        \hat{\mathbf{Y}}^{s,j}_{k|k-1} =& \left[\left\|\mathbf{B}\left(\hat{\bar{\mathbf{x}}}^{j}_{k-1|k-1}\right)\right\|_2,\cdots,\left\|\mathbf{B}\left(\hat{\bar{\mathbf{x}}}^{j}_{k-1|k-1}\right)\right\|_2\right]^T&\text{Scalar model}\\
        \hat{\mathbf{Y}}^{v,j}_{k|k-1} = &\left[\mathbf{B}\left(\hat{\bar{\mathbf{x}}}^{j}_{k-1|k-1}\right)^T,\cdots,\mathbf{B}\left(\hat{\bar{\mathbf{x}}}^{j}_{k-1|k-1}\right)^T\right]^T.&\text{Vector model}
    \end{align}

    For simplicity, we use $\sbullet$ to denote $s$ and $v$. Then we have 
      \begin{align}        \hat{\mathbf{Y}}^{\sbullet}_{k|k-1} =&\sum_{j=0}^{2N}a_j\hat{\mathbf{Y}}^{{\sbullet},j}_{k|k-1}
    \end{align}
and
    \begin{align}
        \mathbf{P}^{\sbullet} &= \sum_{j=0}^{2N} a_j\left(\hat{\mathbf{Y}}^{\sbullet}_{k|k-1}-\hat{\mathbf{Y}}^{^{\sbullet},j}_{k|k-1}\right)\left(\hat{\mathbf{Y}}^{^{\sbullet}}_{k|k-1}-\hat{\mathbf{Y}}^{^{\sbullet},j}_{k|k-1}\right)^T\\
        \mathbf{T}^{^{\sbullet}} &= \sum_{j=0}^{2N} a_j\left(\hat{\mathbf{X}}^j_{k-1|k-1}-\hat{\mathbf{X}}_{k|k-1}\right)\left(\hat{\mathbf{Y}}^{^{\sbullet}}_{k|k-1}-\hat{\mathbf{Y}}^{{\sbullet},j}_{k|k-1}\right)^T.
    \end{align}

    \item Update the state estimation via
    \begin{align}
        K=&\mathbf{T}^{\sbullet}(\mathbf{P}^{\sbullet})^{-1}\\
        \hat{\mathbf{X}}_{k|k}=&\hat{\mathbf{X}}_{k|k-1}+K(\mathbf{Y}^{\sbullet}_{k}-\hat{\mathbf{Y}}^{\sbullet}_{k|k-1})\\
        \Sigma_{k|k}=&\Sigma_{k|k-1}-K\mathbf{P}^{\sbullet}K^T.
    \end{align}
\end{steps}


\section{Fisher information matrix}
To determine the best achievable sensitivity, we perform a Fisher Information Matrix (FIM) analysis~\cite{kay1993fundamentals}.

Before introducing the FIM for scalar measurement and vector measurement, the Jacobian matrix of $\mathbf{B}(\bar{\mathbf{x}}_{i,k})$ with respect to $\bar{\mathbf{x}}_{i,k}$, where $i$ is the sensor index and $k$ is the time step, is given as follows:
\begin{align}
    \mathbf{J}_{i,k}=\frac{3}{r_{i,k}^7}\left(r_{i,k}^2\bar{\mathbf{x}}_{i,k}^T\mathbf{M}\mathbf{I}_3+r_{i,k}^2\bar{\mathbf{x}}_{i,k}\mathbf{M}^T+r_{i,k}^2\mathbf{M}\bar{\mathbf{x}}_{i,k}^T-5\bar{\mathbf{x}}_{i,k}^T\mathbf{M}\bar{\mathbf{x}}_{i,k}\bar{\mathbf{x}}_{i,k}^T\right).
\end{align}
It can be shown that the rank of $\mathbf{J}_{i,k}=3$ if $\bar{\mathbf{x}}_{i,k}\neq[0,0,0]^T$.

\begin{itemize}

\item \textbf{Scalar measurement}:

From~\eqref{mea} we have $y_{i,k}\sim\mathcal{N}(\|\mathbf{B}(\bar{\mathbf{x}}_{i,k})\|_2, \sigma^2)$. Denote the scalar Fisher Information (FI) for the scalar measurement at time $k$ by ${\mathcal{I}}^s_{i,k}$ 
\begin{align}
    \bm{\mathcal{I}}^s_{i,k}=\frac{1}{\sigma^2}\nabla\|\mathbf{B}(\bar{\mathbf{x}}_{i,k})\|_2^T \nabla\|\mathbf{B}(\bar{\mathbf{x}}_{i,k})\|_2
\end{align}
where the $\nabla\|\mathbf{B}(\bar{\mathbf{x}}_{i,k})\|_2$ is the gradient of $\|\mathbf{B}(\bar{\mathbf{x}}_{i,k})\|_2$ with respect to $\bar{\mathbf{x}}_{i,k}$ and given by
\begin{align}
   \nabla\|\mathbf{B}(\bar{\mathbf{x}}_{i,k})\|_2 = \frac{\mathbf{B}(\bar{\mathbf{x}}_{i,k})^T\mathbf{J}_{i,k}}{\|\mathbf{B}(\bar{\mathbf{x}}_{i,k})\|_2}.
\end{align}

\item \textbf{Vector measurement}:

For the vector case, from~\eqref{meas_model_vector}, we can see that $\mathbf{y}_{i,k}\sim\mathcal{N}(\mathbf{g}(\bar{\mathbf{x}}_{i,k}), \bm{\Sigma})$. Denote the Fisher Information Matrix (FIM) for the vector measurement at time $k$ by $\bm{\mathcal{I}}^v_{i,k}$. Then we have
\begin{align}
   \bm{\mathcal{I}}^v_{i,k}=\frac{1}{\sigma^2}\mathbf{J}_{i,k}^T\mathbf{J}_{i,k}.
\end{align}

\end{itemize}

Since measurements are obtained by each sensor independently, the total FIM, $\bm{\mathcal{I}}^s_{k}$ or $\bm{\mathcal{I}}^v_{k}$, for a scalar or vector networks is the summation of the FIMs of all sensors and 
can be written as 
\begin{align}
\bm{\mathcal{I}}^s_{k}=\sum_{i=1}^{n}\bm{\mathcal{I}}^s_{i,k}\;\;\;\text{or}\;\;\;\bm{\mathcal{I}}^v_{k}=\sum_{i=1}^{n}\bm{\mathcal{I}}^v_{i,k}.
\end{align}

Intuitively, one scalar sensor cannot provide sufficient information in estimating a unique location of the target, which can be seen from the FIM corresponding to the scalar measurement model. From~\cite{jauffret2007observability}, the FIM is non-singular if and only if the underlying parameters are (locally) observable. Therefore, $\nabla\|\mathbf{B}(\bar{\mathbf{x}}_{i,k})\|_2$ is a row vector and then  $\text{Rank}(\nabla\|\mathbf{B}(\bar{\mathbf{x}}_{i,k})\|_2^T \nabla\|\mathbf{B}(\bar{\mathbf{x}}_{i,k})\|_2)=1$. On the other hand, 
 $\text{Rank}(\bm{\mathcal{I}}^v_k)=3$ (since $\mathbf{J}_{i,k}^T$ is full rank). As a result, the single sensor with vector measurement is sufficient to provide a unique estimate of the location of the target. This is one of the major advantages of vector magnetometers providing magnetic field measurements, and leads directly to the resilience and tracking sensitivity improvements that we find below. 

The Cram\'er–Rao bound (CRLB) is the best achievable performance bound on the variance of all unbiased estimators for the underlying measurement model and is widely used for benchmarking. Mathematically, the CRLB is the inverse of the FIM. Since the FIM for the single scalar measurement is singular, then its  CRLB will be infinity. However, an array of sensors, at least $3$, with scalar measurement can be used to provide sufficient information and therefore the FIM will be full rank and the CRLB exists.  This is shown in the following simulation results. For efficient interpretation and comparison, the squared root of the trace of the CRLB, $\sqrt{\text{Tr}(CRLB)}$, is used as a measure of the total standard deviation (std)~\cite{rui2014elliptic} in the following analysis.

\begin{figure}[htp!]
         \centering
         \includegraphics[width=0.6\textwidth]{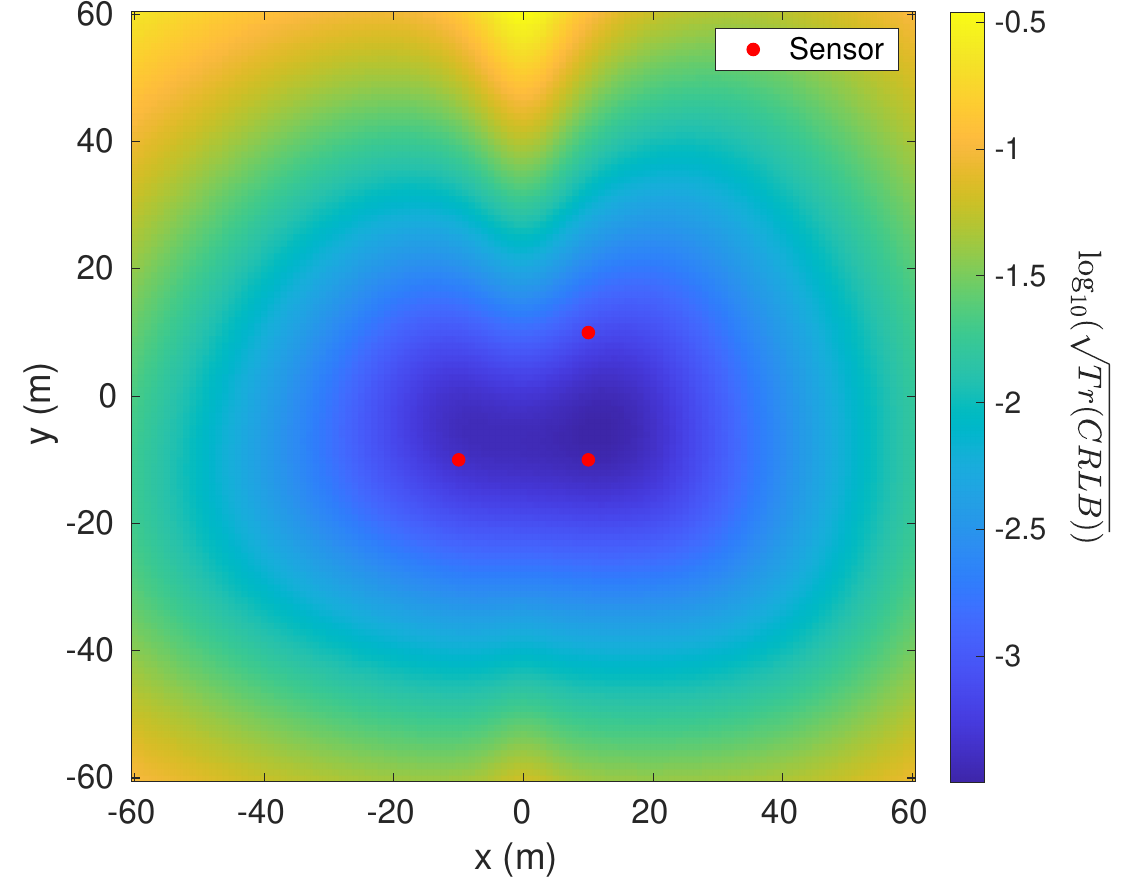}
         \caption{Calculated $\sqrt{\text{Tr}(CRLB)}$ with three sensors (scalar measurement) located at $[10, 10, -25]$~m, $[10, -10, -25]$~m and $[-10, -10, -25]$~m as well as $M=[600, 0, 0]^T$~Am. The results are plotted in $\log_{10}$ scale.}
         \label{fig:crlb_demo_scalar_1}
\end{figure}

\begin{figure}[htp!]
     \centering
     \begin{subfigure}[b]{0.5\textwidth}
         \centering
         \includegraphics[ width=1\textwidth]{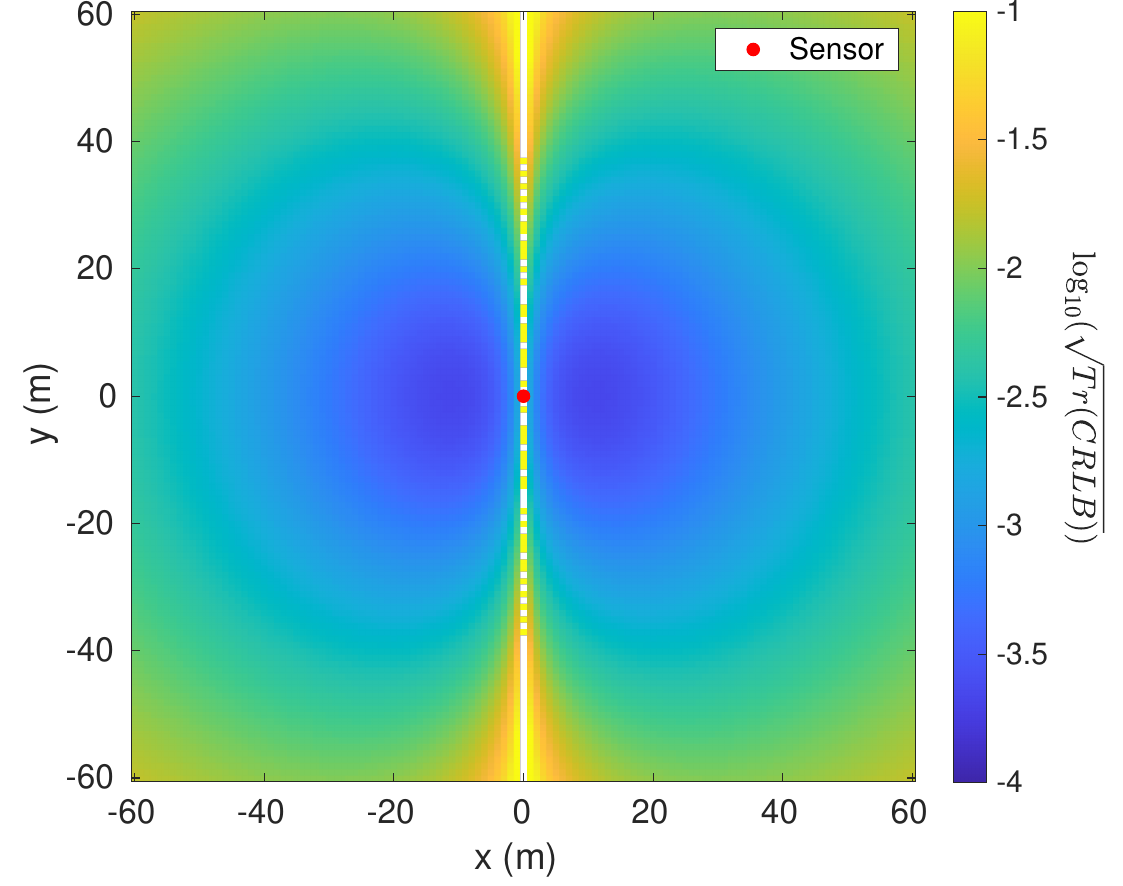}
         \caption{}
         \label{fig:crlb_demo_vector_1}
     \end{subfigure}
    \hspace{-.2cm}
     \begin{subfigure}[b]{0.5\textwidth}
         \centering
         \includegraphics[ width=1\textwidth]{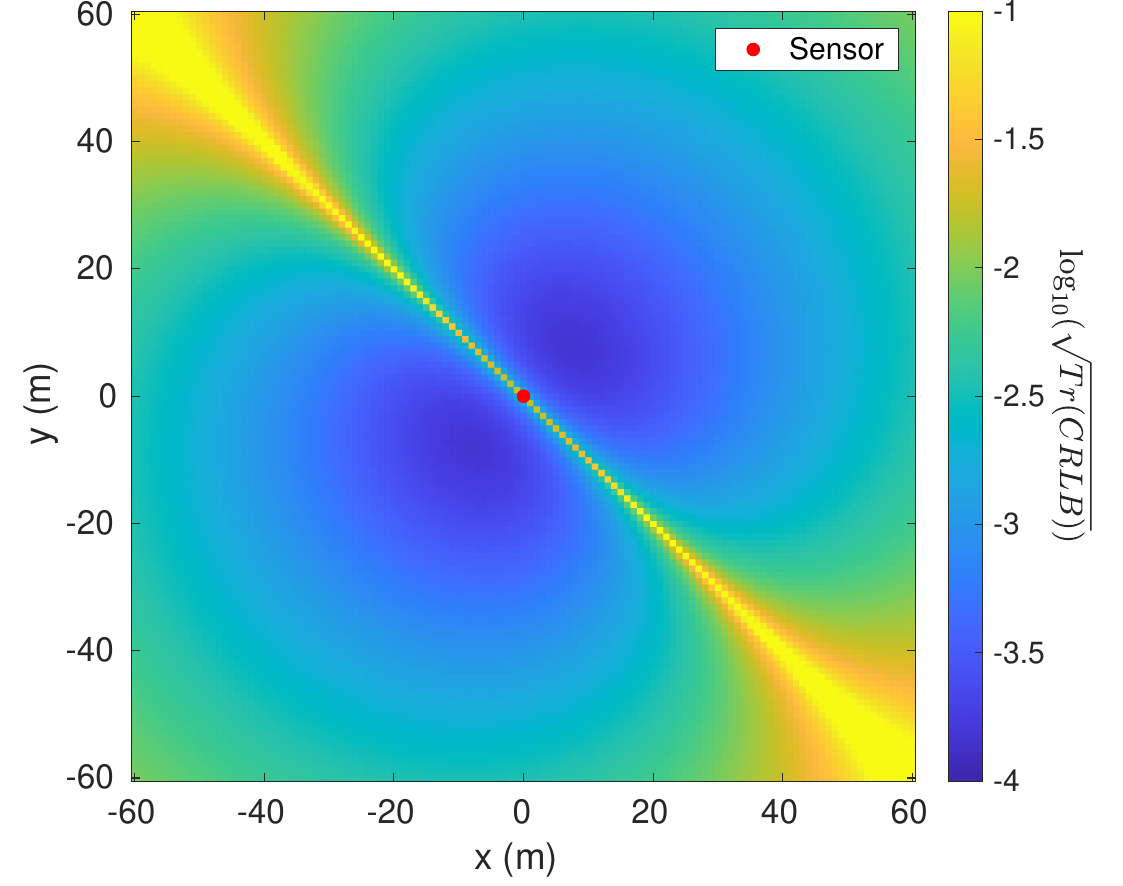}
         \caption{}
         \label{fig:crlb_demo_vector_2}
     \end{subfigure}
        \caption{Calculated $\log_{10}\left(\sqrt{\text{Tr}(CRLB)}\right)$ with 3 sensor (vector measurement) located at $[0, 0, -25]$~m and different $M$. (a) $M=[600, 0, 0]^T$~Am; (b) $M=[600, 600, 2]^T$~Am}
        \label{fig:crlb_demo_vector}
\end{figure}

Fig.~\ref{fig:crlb_demo_scalar_1} shows an example of squared root of trace of CRLB, i.e. $\sqrt{\text{Tr}(CRLB)}$, with 3 sensors (scalar measurement) located at $[10, 10, -24]$~m, $[10, -10, -24]$~m and $[-10, -10, -24]$~m, and $M=[600, 0, 0]^T$~A$\cdot$m$^2$. We can see that 3 sensors are enough to provide sufficient information in estimating the target because of the non-singular FIM.
On the other hand, Fig.~\ref{fig:crlb_demo_vector} shows two examples of trace of CRLB with 1 sensor (vector measurement) located at $[0, 0, -24]$~m and different $M$. It can be seen that, in Fig.~\ref{fig:crlb_demo_vector_1}, the trace of CRLB is symmetric about the $x$-axis while, in Fig.~\ref{fig:crlb_demo_vector_2}, it is symmetric about the $y=x$ line roughly, which means that the value of $M$ impacts the symmetricality of $\sqrt{\text{Tr}(CRLB)}$. 

Another notable phenomenon in Fig.~\ref{fig:crlb_demo_vector_1} is that the target is unobservable, labeled by white points, if it lies on the $x$-axis due to the FIM being singular since $M=[600, 0, 0]^T$~Am. However, if $M$ is non-zero vector and cannot be scaled to all-one vector, there will be no unobservable point.

\section{Performance comparison of scalar and vector arrays}

In this section, we consider two scenarios to demonstrate the relative performance of the networks with different measurement models, i.e. scalar and vector. In the first scenario, by changing the measurement model, the position of the dipole ($z$-axis), the CRLB of the two models are demonstrated and plotted to show that the vector measurement outperforms the scalar one. In the second scenario, a practical example is considered to show how the performance of the sensor array by changing the number and spacing of sensors as well as the target's trajectories.

\subsection{Scenario I}

In this scenario, $49$ sensors are placed in a square grid over area $[-400, 800]~\text{m}\times[400, -800]~\text{m}$.  The target is assumed to be on the surface (i.e. $z=0$), with the sensor array depth either at $z=-25$~m or $z=-80$~m. In the following simulations of CRLB, the std of measurement noise is $10$~pT. Since the CRLB is proportional to the std of measurement noise, similar results can be derived for other noise levels. The ground truth motion of the target is circular above the sensor array and plotted with red dashed line. 

\begin{figure}[htp!]
\centering
 \begin{subfigure}{0.5\textwidth}
         \includegraphics[width=1\textwidth]{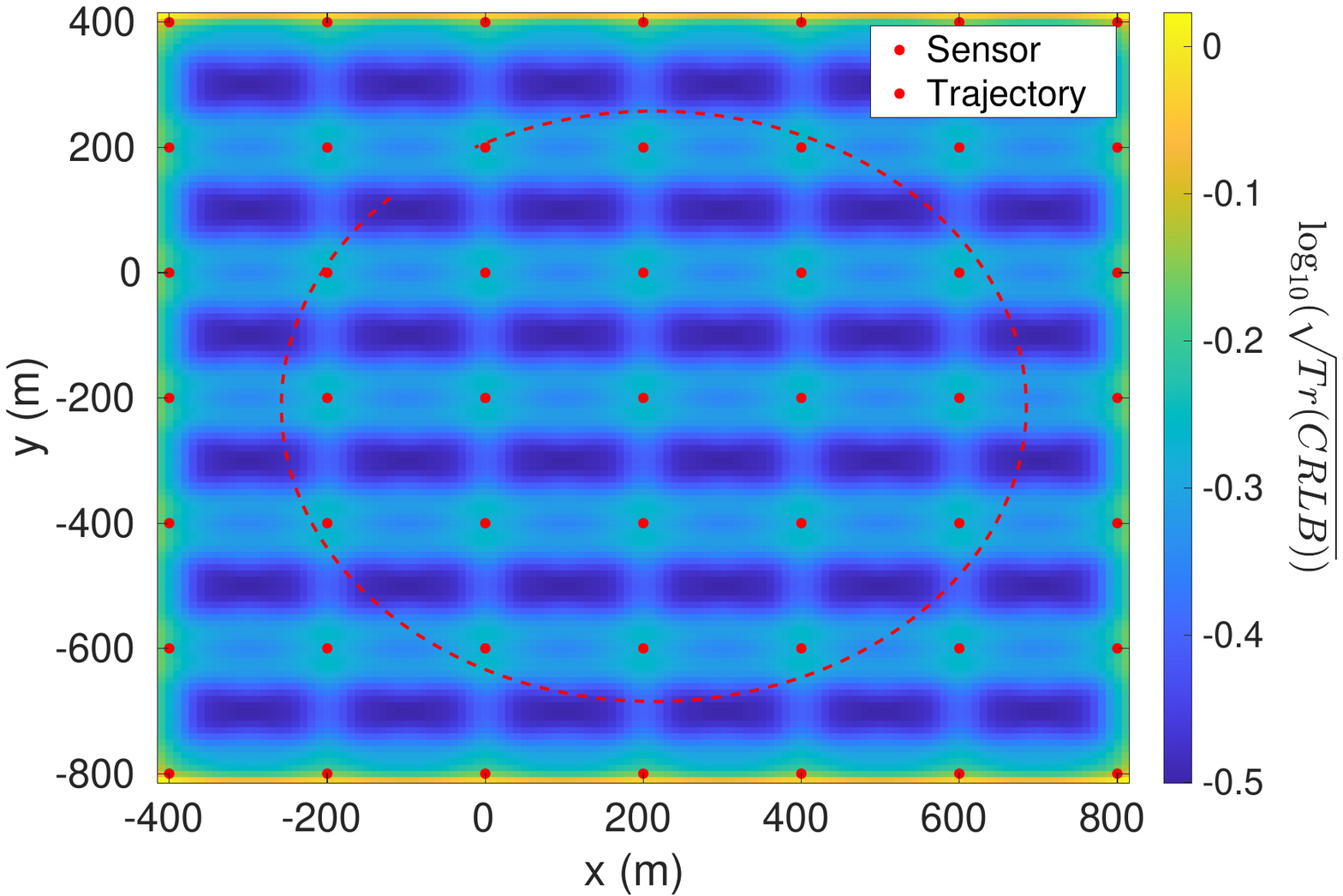}
         \caption{}
         \label{fig:CRLB1_1}
 \end{subfigure}
\hspace{-.2cm}
\begin{subfigure}{0.5\textwidth}
         \includegraphics[width=1\textwidth]{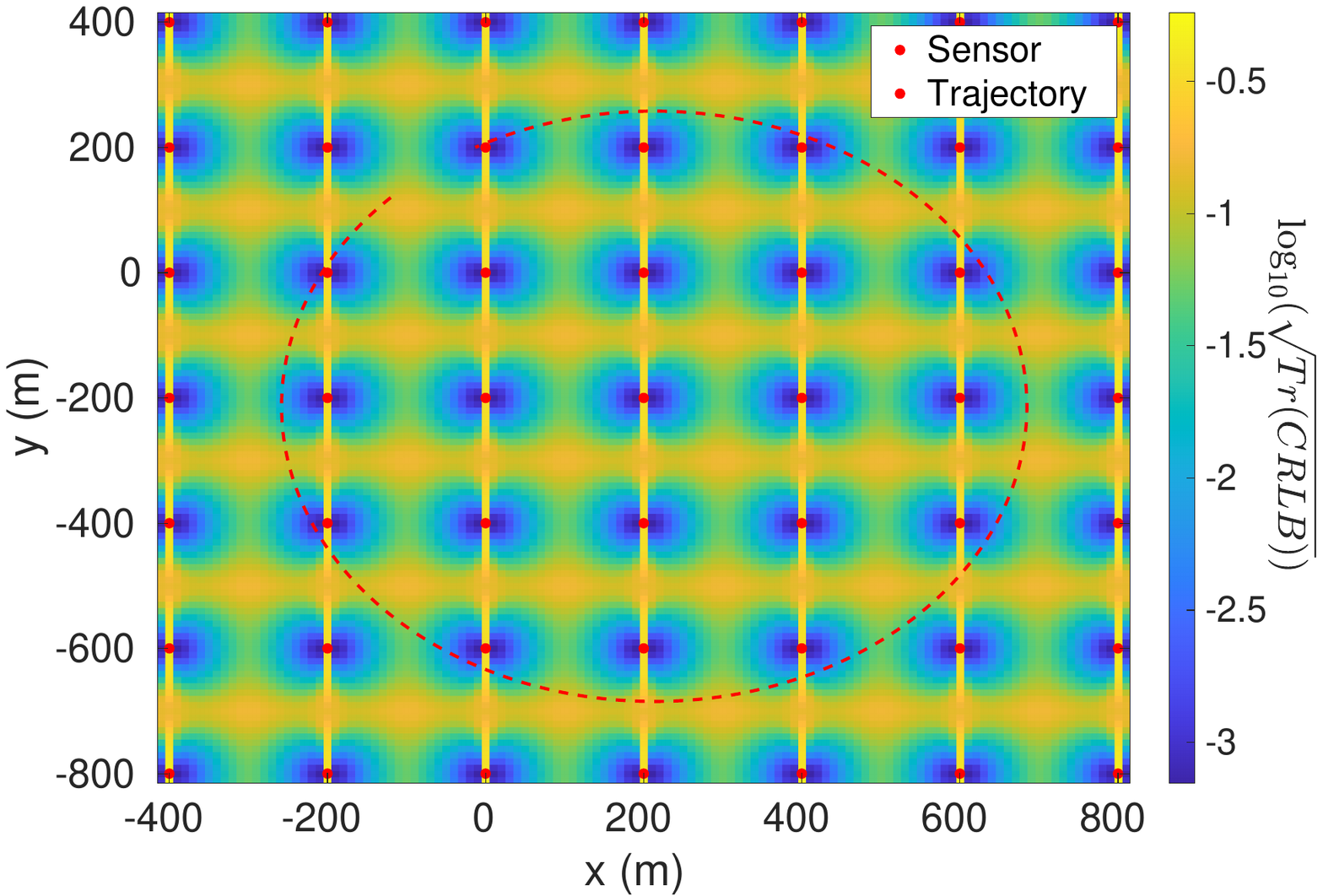}
         \caption{}
         \label{fig:CRLB1_2}
 \end{subfigure}
 
 \medskip
 
  \begin{subfigure}{0.49\textwidth}
         \includegraphics[width=1\textwidth]{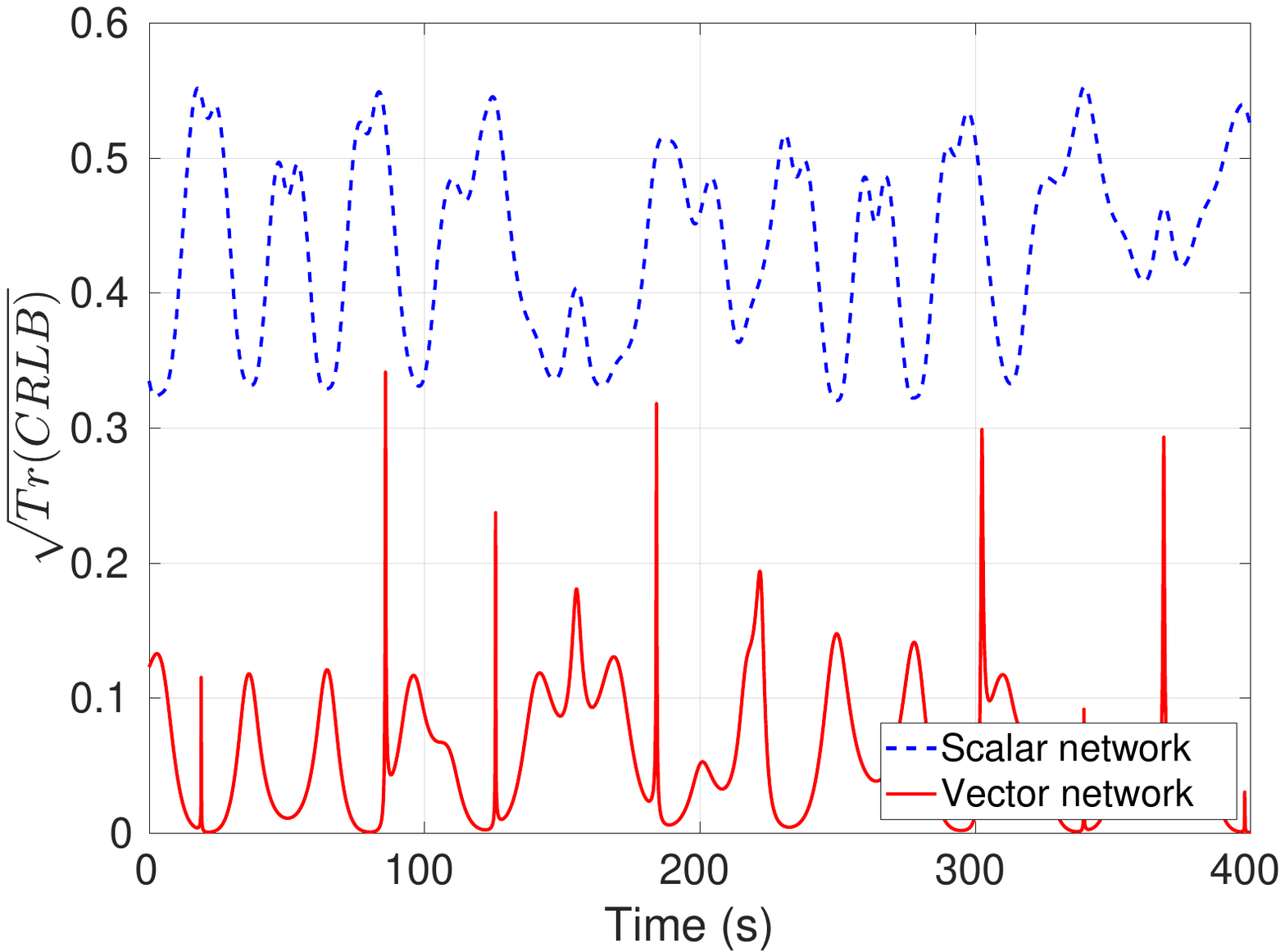}
         \caption{}
         \label{fig:CRLB1_3}
 \end{subfigure}

 \caption{The plot of $\log_{10}\left(\sqrt{\text{Tr}(CRLB)}\right)$ over the interested area and true trajectory.  The std of noise is $10$~pT and sensors' depth is fixed at $z=-25$~m. (a) scalar model; (b) vector model; (c) the $\sqrt{\text{Tr}(CRLB)}$ along the trajectory.}
        \label{fig:CRLB1}
\end{figure}

In Fig.~\ref{fig:CRLB1_1} and~\ref{fig:CRLB1_2}, the $\sqrt{\text{Tr}(CRLB)}$ of the interested area for scalar and vector measurement models are plotted respectively. In these two figures, the height of sensors are fixed to be $-25$m. In Fig.~\ref{fig:CRLB1_3}, the $\sqrt{\text{Tr}(CRLB)}$ along the trajectory are plotted for scalar and vector measurement models. It can be seen that the latter one outperforms the former one in terms of the CRLB. 

\begin{figure}[htp!]
     \centering
          \begin{subfigure}[b]{0.5\textwidth}
         \centering
         \includegraphics[width=1\textwidth]{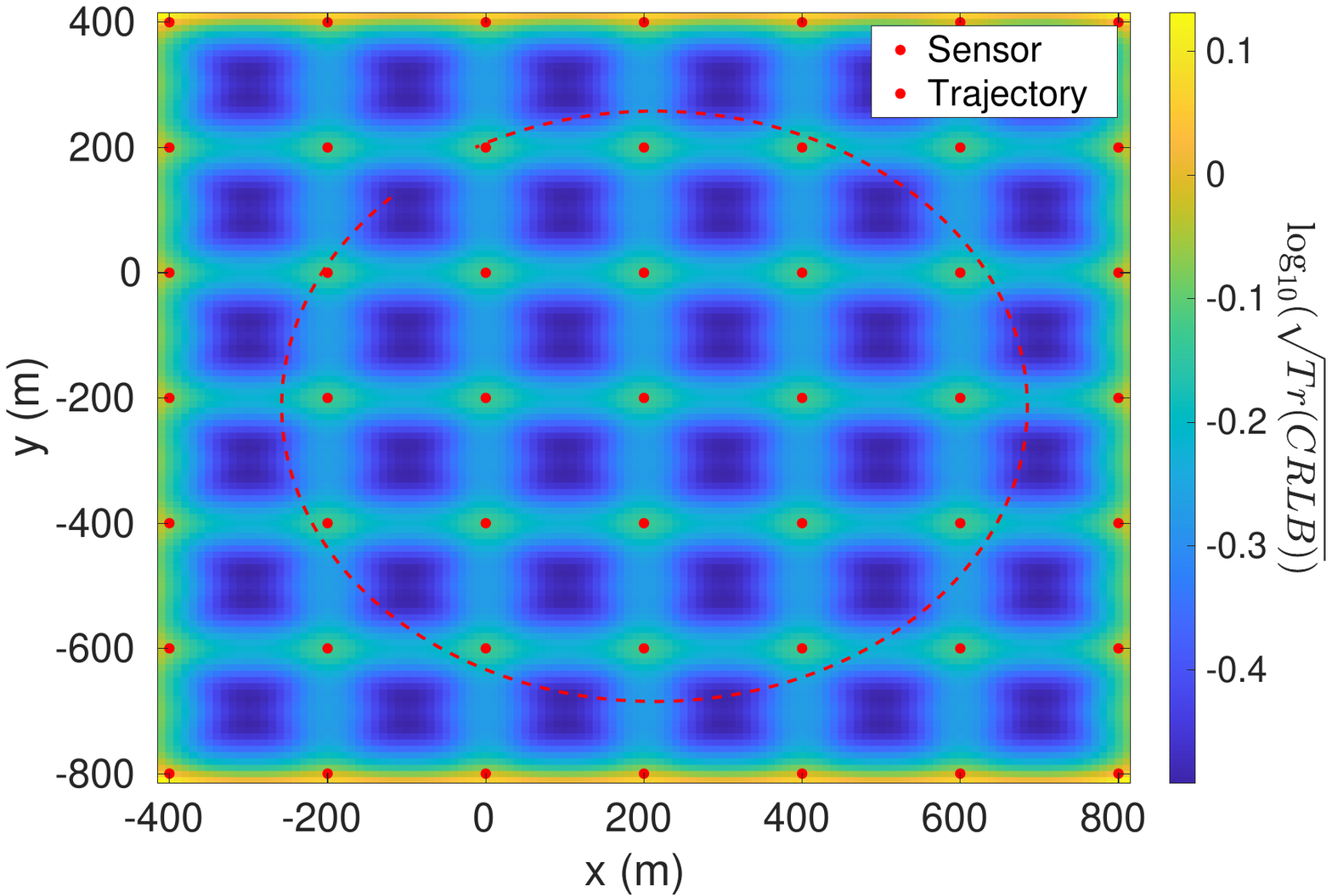}
         \caption{}
         \label{fig:CRLB2_1}
     \end{subfigure}
\hspace{-.2cm}
\begin{subfigure}[b]{0.5\textwidth}
         \centering
         \includegraphics[width=1\textwidth]{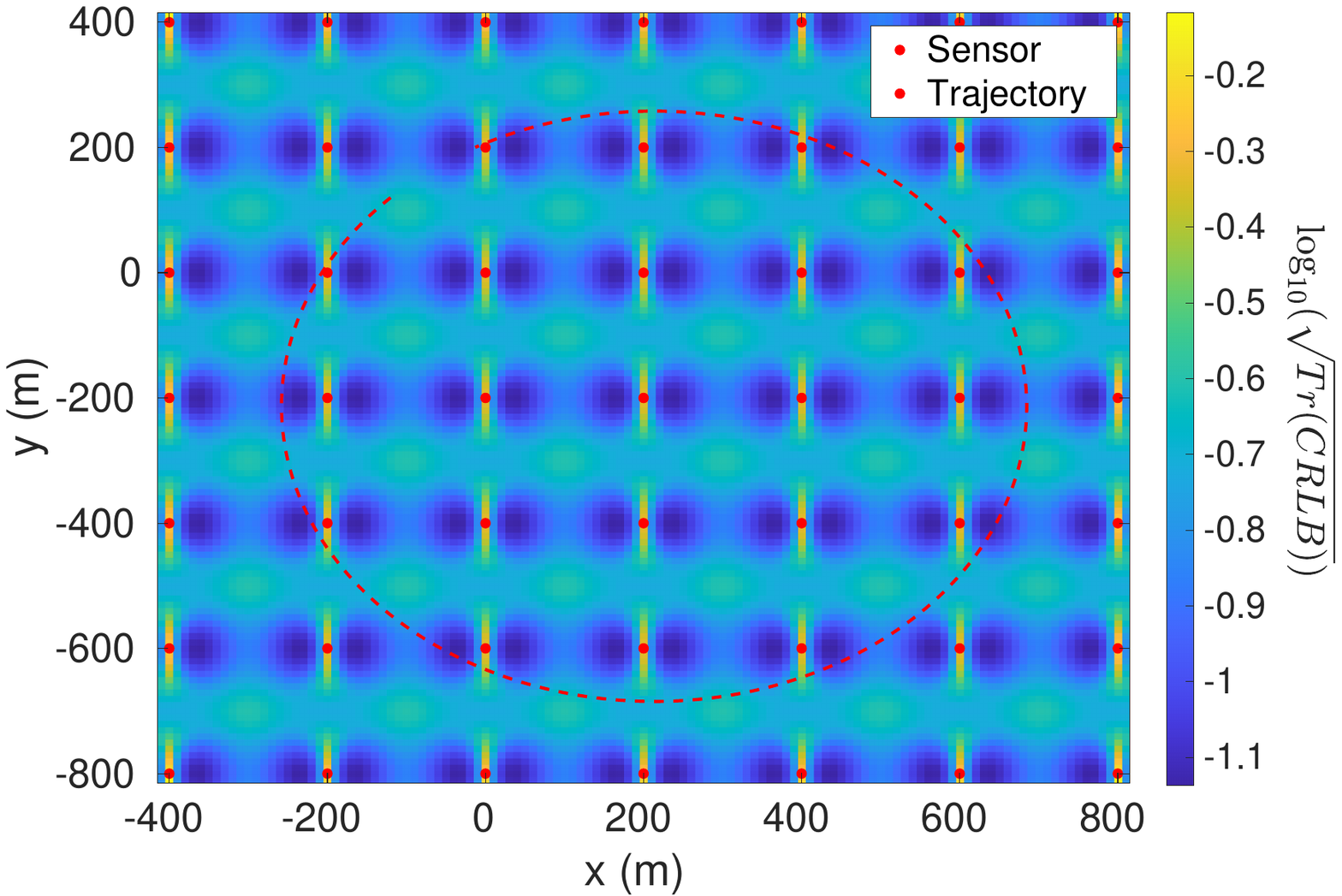}
         \caption{}
         \label{fig:CRLB2_2}
     \end{subfigure}
     
      \medskip
      
    \begin{subfigure}[b]{0.49\textwidth}
         \centering
         \includegraphics[width=1\textwidth]{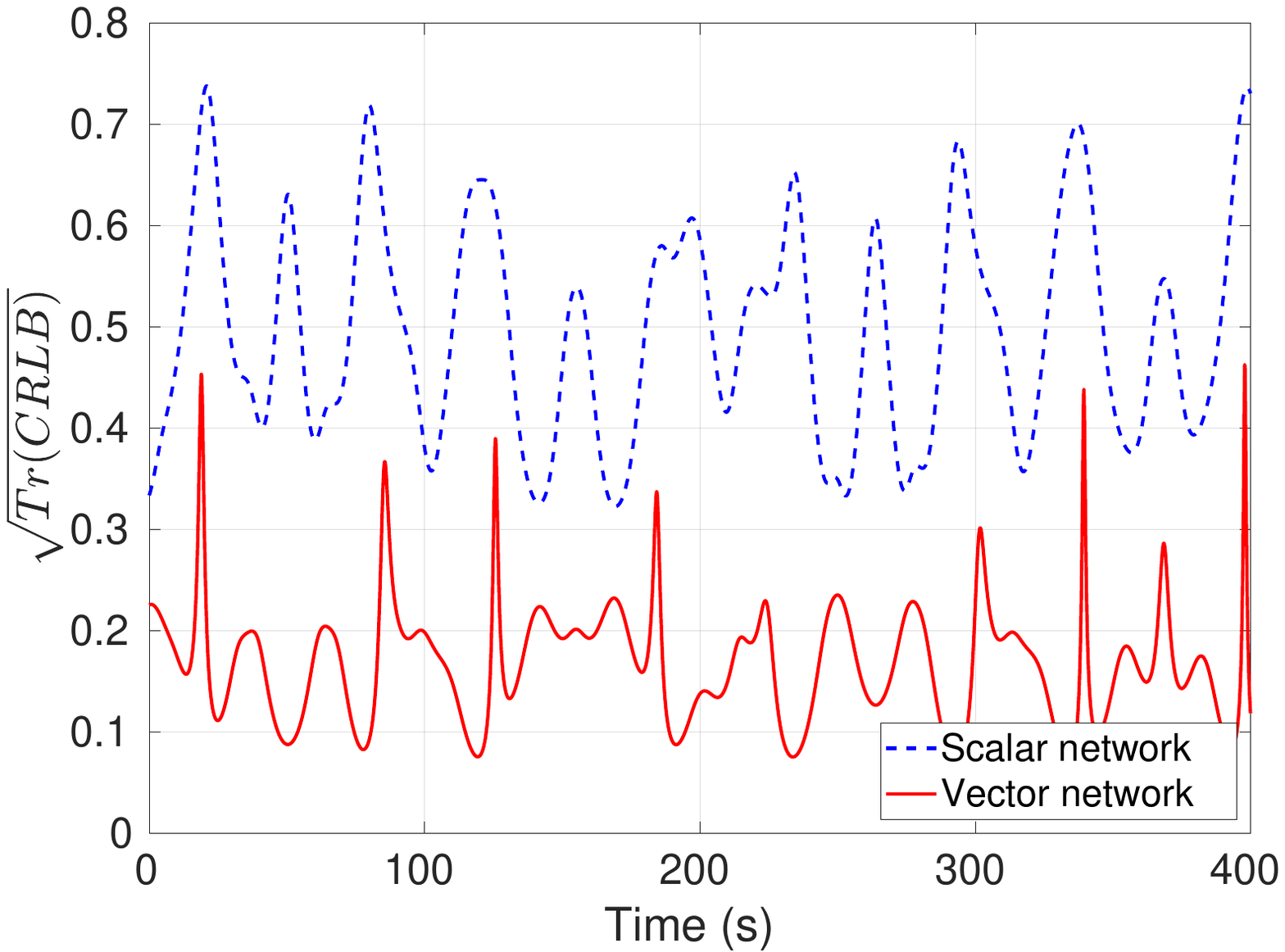}
         \caption{}
         \label{fig:CRLB2_3}
     \end{subfigure}
        \caption{The plot of $\log_{10}\left(\sqrt{\text{Tr}(CRLB)}\right)$ over the interested area and true trajectory.  The std of noise is $10$~pT and sensors' heights are fixed to $-80$~m. (a) scalar model; (b) vector model; (c) the $\sqrt{\text{Tr}(CRLB)}$ along the trajectory.}
        \label{fig:CRLB2}
\end{figure}

Another example to see the improvement of the vector magnetometer array is for increasing depth.  Fig.~\ref{fig:CRLB2_1} and Fig.~\ref{fig:CRLB2_2} compare the scalar and vector arrays at a depth of $z=-80$~m, i.e., over three times deeper than the case in Fig.~\ref{fig:CRLB1}, with same noise level. Similarly, the $\sqrt{\text{Tr}(CRLB)}$ along the trajectory are plotted in Fig.~\ref{fig:CRLB2_3} for scalar and vector measurement models. 

From these figures, it can be seen that the vector measurement model outperforms the scalar one in theoretical results, i.e. CRLB. 


\subsection{Scenario II}\label{sec:sencario_2}
\subsubsection{Part I}\label{sec:sencario_2_1}

\begin{figure}[htp!]
     \centering
         \includegraphics[width=0.6\textwidth]{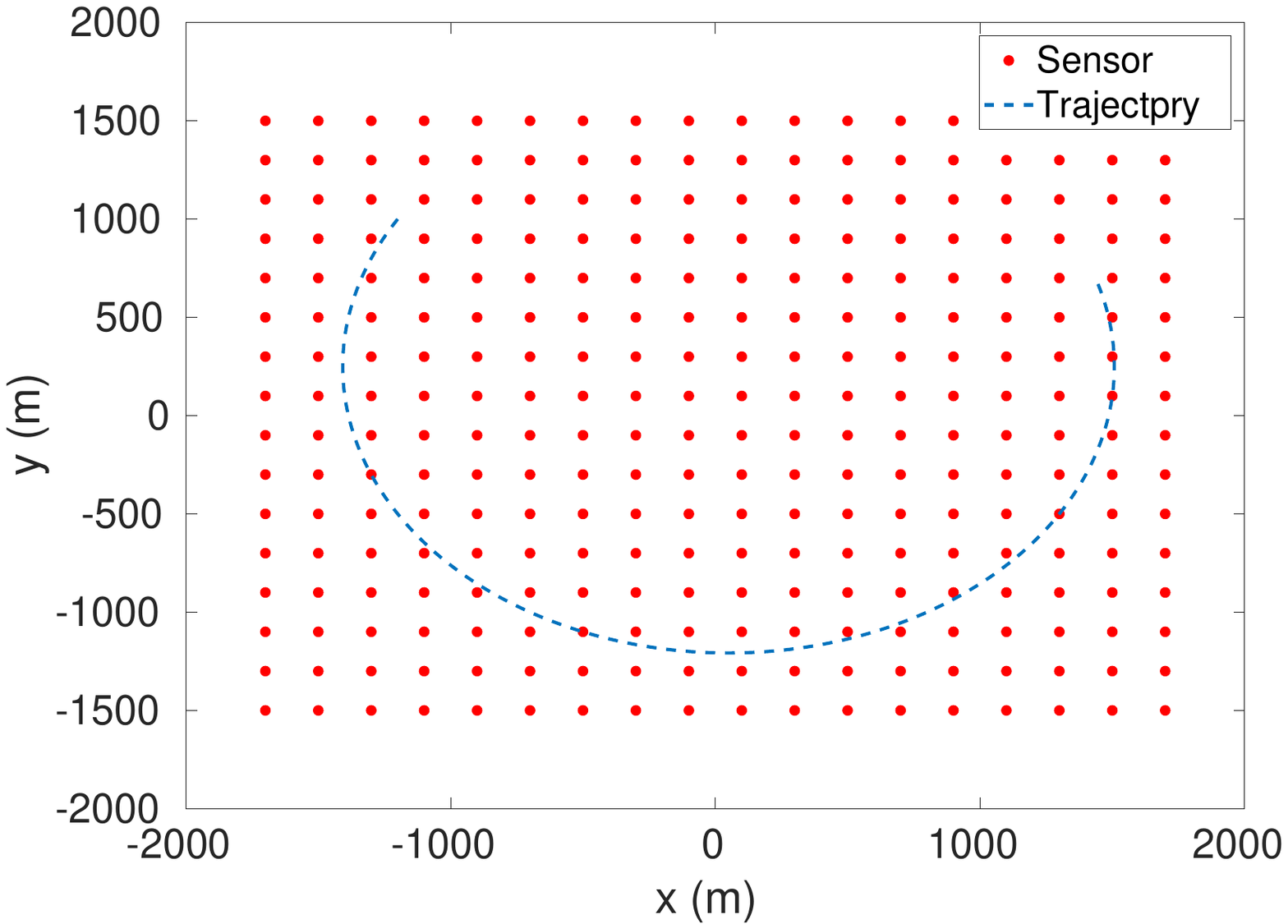}
        \caption{Diagram showing the sensor network superimposed to highlight one of our proposed use cases. We assume an $m \times n$ sensor array with $d$ spacing along the x-axis and y-axis respectively covering an area of ~$3500 \times 3000$ m$^2$. In this particular example showing in the figure, there are $288$ sensors with spacing $200$~m. The red dots are sensors while the blue curve is the trajectory of the object to be tracked.}
        \label{fig:PortPhilipBayMap}
\end{figure}

In this scenario, we assume that the sensors are placed on a seabed and the target is moving circularly with a speed \textasciitilde $11$~knots on the sea surface, see Fig.~\ref{fig:PortPhilipBayMap} for illustration. We also assume in the following simulations that the depth of sensors is $24$~m which is the approximate average depth of Port Phillip Bay seabed~\cite{holdgate2001marine}. In order to investigate the impact the of the density of sensor array, the number of sensors will be varied according to the spacing distance, i.e. less spacing distance corresponds to more sensors. In particular, the sensor spacings are set to be $200$~m, $300$~m and $400$~m in the simulations, corresponding to the number of sensors as $288$, $132$ and $72$. In addition, the std of the measurement noise is set to be $32$~pT, $158$~pT and $320$~pT, and the magnetic moment of the target is $[600,0, 0]^T$ A$\cdot$m$^2$. Finally, the total running time of the trajectory is $~16.6$~min with $500$ time Monte Carlo simulations.

\begin{figure}[htp!]
 \begin{subfigure}{0.49\textwidth}
     \includegraphics[ width=1\textwidth]{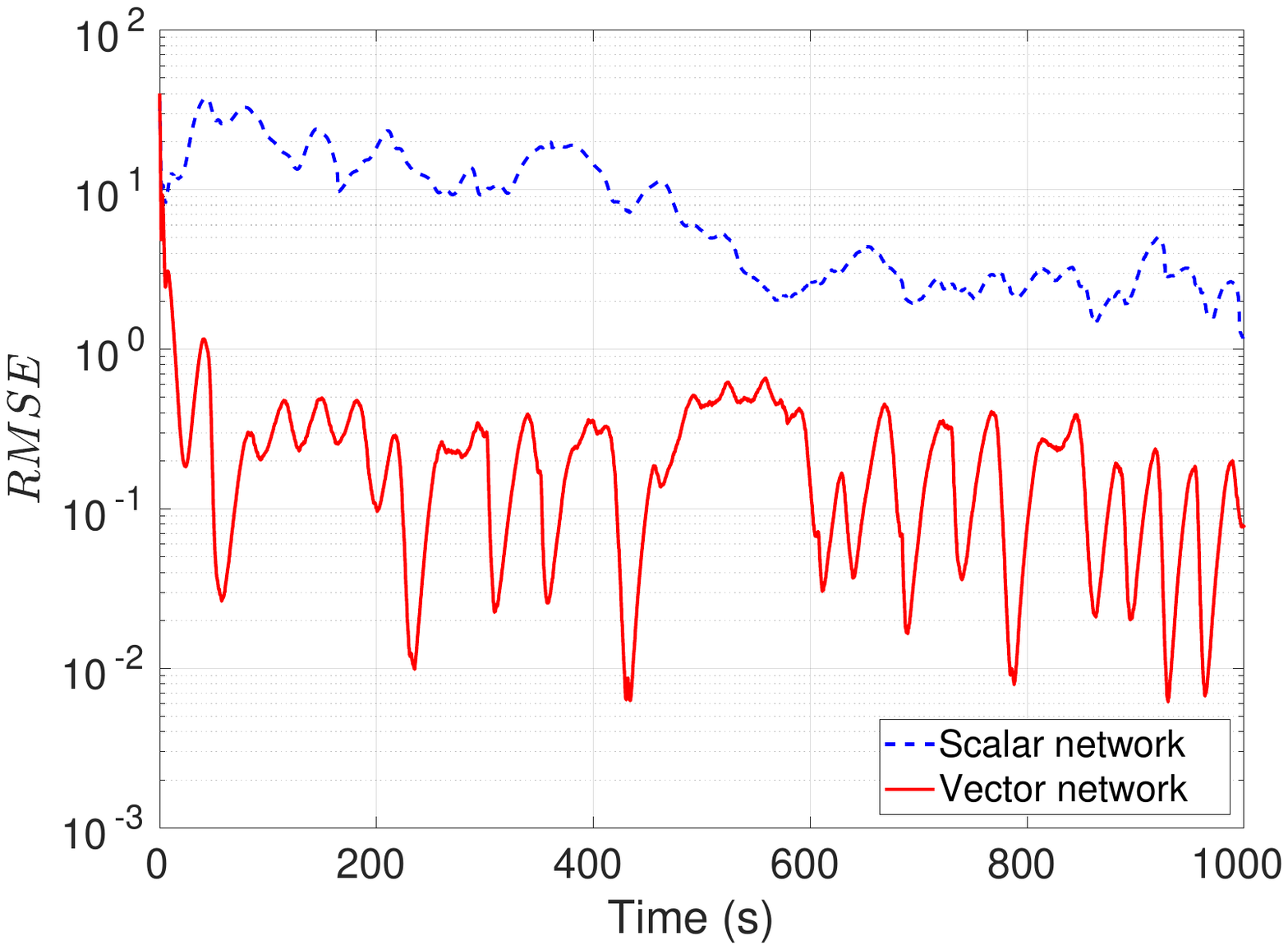}
         \caption{}
         \label{fig:real_sensor_s1_1}
 \end{subfigure}
 \hfill
 \begin{subfigure}{0.49\textwidth}
     \includegraphics[width=1\textwidth]{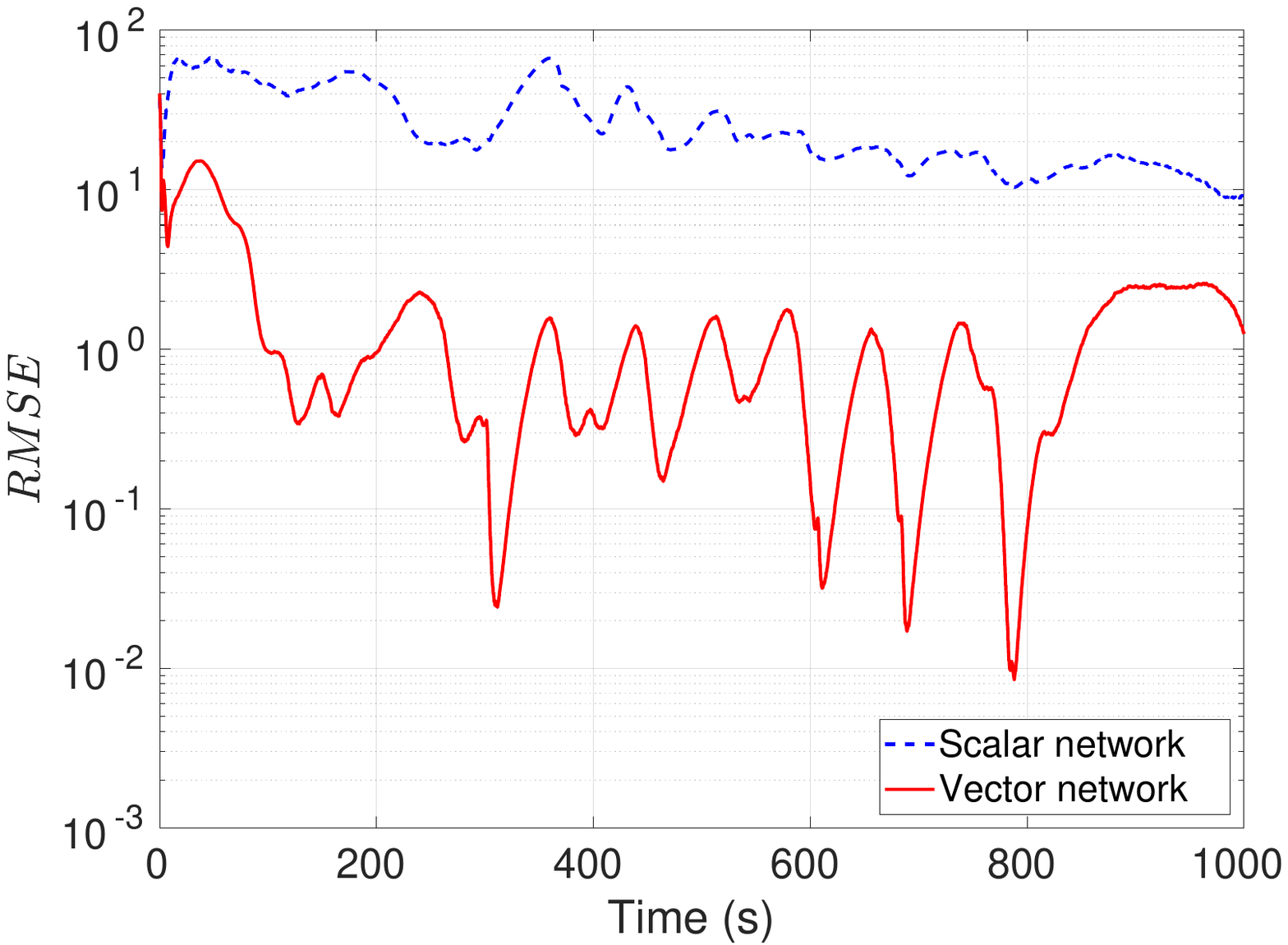}
         \caption{}
         \label{fig:real_sensor_s1_2}
 \end{subfigure}
 
 \medskip
 \begin{subfigure}{0.49\textwidth}
        \includegraphics[ width=1\textwidth]{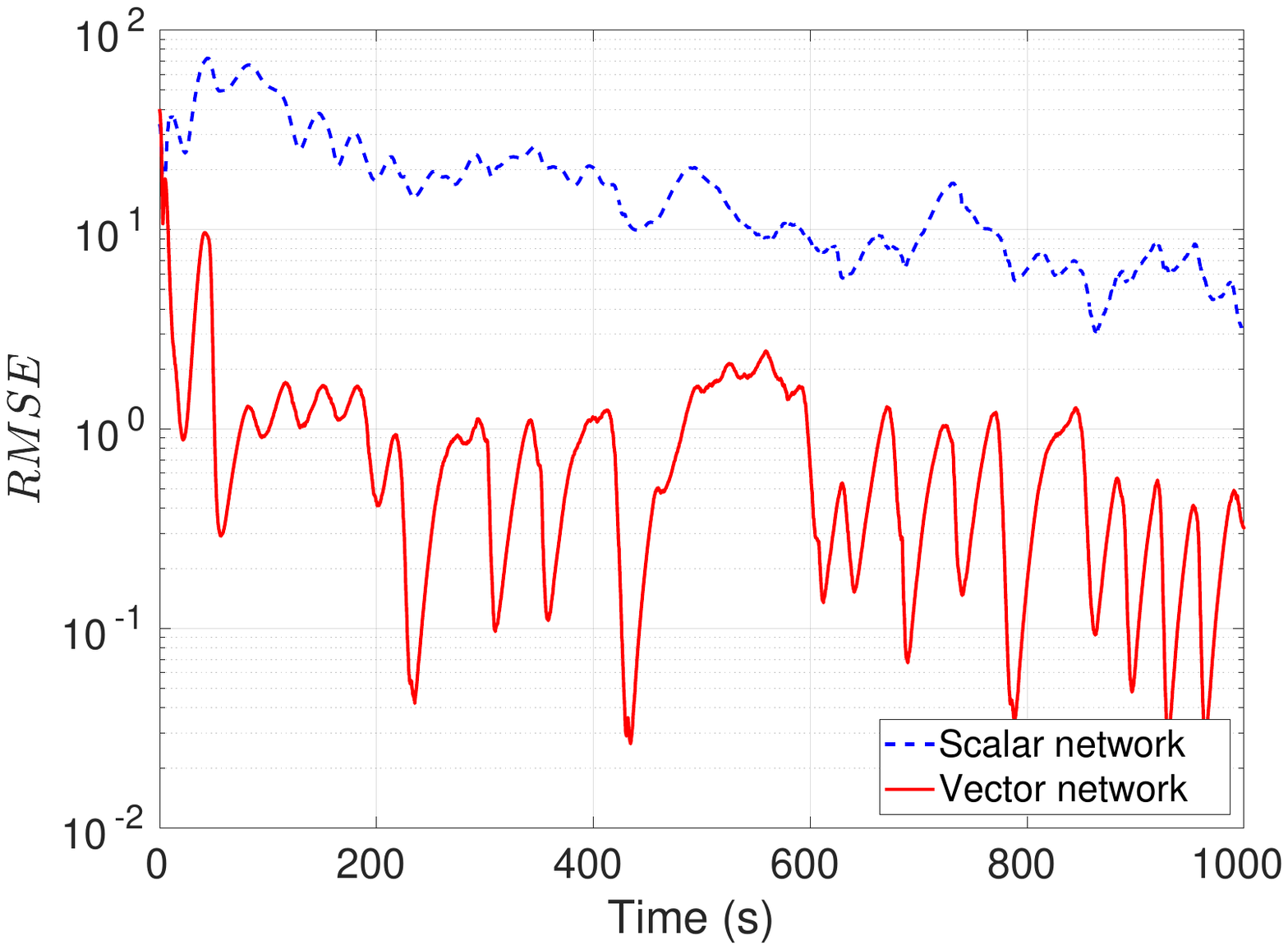}
         \caption{}
         \label{fig:real_sensor_s2_1}
 \end{subfigure}
 \hfill
 \begin{subfigure}{0.49\textwidth}
     \includegraphics[ width=1\textwidth]{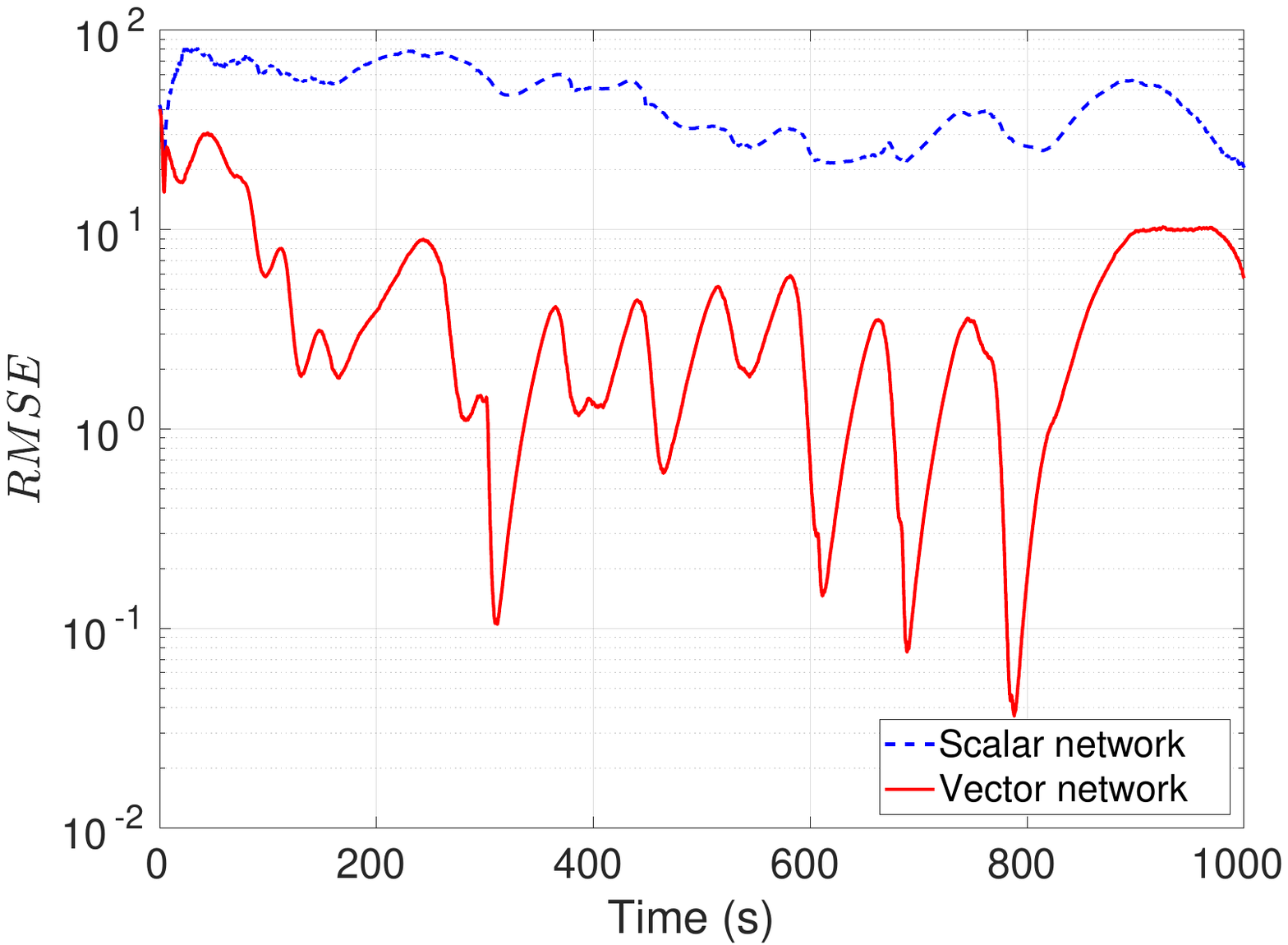}
         \caption{}
         \label{fig:real_sensor_s2_2}
 \end{subfigure}

 \caption{The simulated RMSEs with noise level $160$~pT and different sensor spacings. (a) Sensor spacing is $200$~m with $32$~pT; (b) Sensor spacing is $300$~m with $32$~pT; (c) Sensor spacing is $200$~m with $160$~pT; (d) Sensor spacing is $300$~m with $160$~pT.  In all cases, vector magnetometer arrays significantly outperform scalar magnetometer arrays for target tracking.}
        \label{fig:real_sensor_sim}
\end{figure}

The simulation results with std of noise $32$~pT and $160$~pT as well as sensor spacing $200$~m and $300$~m are shown in Fig.~\ref{fig:real_sensor_sim}. 

The other cases are not plotted as the failure percentages, percentages of divergence of tracking or tracking error is greater than $200$~m, of the scalar vector is great so that the comparison is not meaningful, see Table~\ref{failure_percentage_real}.

\begin{table}[H]
\centering
\begin{tabular}{|c|c|c|c|}
\hline
 Noise level            & Sensor spacing & Scalar network & Vector network  \\ \hline
\multirow{3}{*}{$32$~pT} & $200$     & $8.3\%$ & $0\%$  \\ \cline{2-4} 
                        & $300$     & $19.7\%$ & $0\%$  \\ \cline{2-4} 
                        & $400$     & $26.0\%$ &  $0.3\%$ \\ \hline
\multirow{3}{*}{$160$~pT} & $200$     & $21.0\%$ & $0\%$  \\ \cline{2-4} 
                        & $300$     & $51.0\%$ & $0\%$  \\ \cline{2-4} 
                        & $400$     & $84.7\%$ & $0.3\%$  \\ \hline
\multirow{3}{*}{$320$~pT} & $200$     & $35.6\%$ & $0.5\%$  \\ \cline{2-4} 
                        & $300$     & $76.7\%$ & $8\%$  \\ \cline{2-4} 
                        & $400$     & $93.7\%$ & $12.2\%$  \\ \hline
\end{tabular}
\caption{The failure percentages of the simulations with different noise levels and sensor spacings.  Here we see that vector networks provide significantly enhanced tracking resilience.}
\label{failure_percentage_real}
\end{table}

From the Fig.~\ref{fig:real_sensor_sim} and Table~\ref{failure_percentage_real}, we can see that the vector network outperforms the scalar one in terms of accuracy and successful tracking performance. In the figures, one can notice that the minimal RMSE that the vector model can achieve is as low as $0.001$~m comparing to the scalar model $0.3$~m. It also can be seen that, in the Fig.~\ref{fig:real_sensor_s1_1}, the RMSE of vector model outperforms at least (approximately) $3$ times as many as scalar model, at $500$~second point, while at most (approximately) $15$ times, at $780$~second point. The similar results can be observed from other figures. 

\subsubsection{Part II}

The undersea domain can be harsh, and replacing sensors may be difficult.  Therefore it is important to consider the consequences of permanent sensor outages, to determine if the sensor arrays are able to fail gracefully.

In this scenario, we therefore consider the initial grid networks, and fail a randomly chosen subset of the sensors.  By providing the RMSE tracking fidelity over the full trajectory and Monte-Carlo averaging over the failed detectors, we are able to get a statistical measure for the resilience of the network. 

Our results show significantly improved resilience for the vector network over the scalar network, highlighting the benefit of scalar magnetometer networks for resilient monitoring networks. In the simulations, three cases are considered, i.e. $10$ sensors failed (Case I); $15$ sensors failed (Case II); and $20$ sensors failed (Case III).  Table~\ref{failure_percentage} shows the percentage of tracking failures as a function of increasing failed sensors. As can be seen, at these sensitivities, the vector network maintains tracking even with 20 failed sensors out of the 288 in the original grid.  In this limit, the scalar network fails to track at 25.9~\% of cases.  This highlights the improved robustness of the vector network relative to scalar magnetometer networks.

\begin{table}[H]
\centering
\begin{tabular}{|c|c|c|c|}
\hline
                        & Noise level & Scalar network & Vector network  \\ \hline
\multirow{2}{*}{Case I} & $32$~pT     & $9.3\%$ & $0\%$  \\ \cline{2-4} 
                        & $160$~pT     & $23.5\%$ & $0\%$  \\ \hline
\multirow{2}{*}{Case II} & $32$~pT     & $10.1\%$ & $0\%$  \\ \cline{2-4} 
                        & $160$~pT     & $24.1\%$ & $0\%$  \\ \hline     
\multirow{2}{*}{Case II} & $32$~pT     & $12.4\%$ & $0\%$  \\ \cline{2-4} 
                        & $160$~pT     & $25.9\%$ & $0\%$  \\ \hline                           
\end{tabular}
\caption{The failure percentages of the cases}
\label{failure_percentage}
\end{table}

\section{Conclusions}

Monitoring of commercial and non-commercial sea traffic is becoming increasingly important for maritime safety.  Magnetometer arrays have the potential to provide crucial information for monitoring traffic and detecting threats.  Our analysis highlights the added advantage of vector magnetometers over scalar magnetometers.  

Vector arrays typically provide more than a three-fold improvement in performance due to their enhanced ability to localise targets and are more resilient to the loss of sensors than comparable scalar arrays.  This is a strong motivating factor for exploring practical vector magnetometer solutions suitably for the undersea domain, such as diamond in fibre or other ruggedised diamond-based solutions for magnetometry.

Although tracking accuracy has been greatly improved in three-axis sensor networks, the deployment of vector sensors with ignorable errors underwater is challenging and we will continue our research in this aspect. 
\bibliographystyle{plain}
\bibliography{MagNetwork}

\end{document}